\documentclass[twocolumn,showpacs,aps,prl,superscriptaddress,floatfix]{revtex4}
\usepackage{graphicx}
\usepackage{dcolumn}
\usepackage{multirow}
\usepackage{amsmath}
\usepackage{epsfig}
\usepackage{color}

\RequirePackage{xspace}

\newcommand{\BABARPubYear}    {08}
\newcommand{\BABARPubNumber} {038}
\newcommand{\SLACPubNumber}{13343}

\newcommand{\arxiv}{0808.3524v2}

\usepackage{relsize}
\def\babar{\mbox{\slshape B\kern-0.1em{\smaller A}\kern-0.1em
    B\kern-0.1em{\smaller A\kern-0.2em R}}}

\def\epem       {\ensuremath{e^+e^-}\xspace}

\def\qqbar {\ensuremath{q\overline q}\xspace}

\def\Kbar  {\kern 0.2em\overline{\kern -0.2em K}{}\xspace}

\def\Kz    {\ensuremath{K^0}\xspace}
\def\Kzb   {\ensuremath{\Kbar^0}\xspace}
\def\KzKzb {\ensuremath{\Kz \kern -0.16em \Kzb}\xspace}
\def\Kp    {\ensuremath{K^+}\xspace}
\def\Km    {\ensuremath{K^-}\xspace}

\def\KpKm  {\ensuremath{\Kp \kern -0.16em \Km}\xspace}

\def\Dbar    {\kern 0.2em\overline{\kern -0.2em D}{}\xspace}

\def\Dz      {\ensuremath{D^0}\xspace}
\def\Dzb     {\ensuremath{\Dbar^0}\xspace}
\def\DzDzb   {\ensuremath{\Dz {\kern -0.16em \Dzb}}\xspace}
\def\Dp      {\ensuremath{D^+}\xspace}
\def\Dm      {\ensuremath{D^-}\xspace}

\def\DpDm    {\ensuremath{\Dp {\kern -0.16em \Dm}}\xspace}

\def\Bbar    {\kern 0.18em\overline{\kern -0.18em B}{}\xspace}

\def\BB      {\ensuremath{B\Bbar}\xspace} 
\def\Bz      {\ensuremath{B^0}\xspace}
\def\Bzb     {\ensuremath{\Bbar^0}\xspace}
\def\BzBzb   {\ensuremath{\Bz {\kern -0.16em \Bzb}}\xspace}
\def\Bu      {\ensuremath{B^+}\xspace}
\def\Bub     {\ensuremath{B^-}\xspace}

\def\BpBm    {\ensuremath{\Bu {\kern -0.16em \Bub}}\xspace}

\def\BorBbar    {\kern 0.18em\optbar{\kern -0.18em B}{}\xspace}
\def\DorDbar    {\kern 0.18em\optbar{\kern -0.18em D}{}\xspace}
\def\KorKbar    {\kern 0.18em\optbar{\kern -0.18em K}{}\xspace}

\mathchardef\Upsilon="7107
\def\Y#1S{\ensuremath{\Upsilon{(#1S)}}\xspace}

\def\FourS {\Y4S}

\mathchardef\Deltares="7101
\mathchardef\Xi="7104
\mathchardef\Lambda="7103
\mathchardef\Sigma="7106
\mathchardef\Omega="710A

\def\Deltabar{\kern 0.25em\overline{\kern -0.25em \Deltares}{}\xspace}
\def\Lbar{\kern 0.2em\overline{\kern -0.2em\Lambda\kern 0.05em}\kern-0.05em{}\xspace}
\def\Sigbar{\kern 0.2em\overline{\kern -0.2em \Sigma}{}\xspace}
\def\Xibar{\kern 0.2em\overline{\kern -0.2em \Xi}{}\xspace}
\def\Obar{\kern 0.2em\overline{\kern -0.2em \Omega}{}\xspace}
\def\Nbar{\kern 0.2em\overline{\kern -0.2em N}{}\xspace}
\def\Xb{\kern 0.2em\overline{\kern -0.2em X}{}\xspace}

\def\BR         {{\ensuremath{\cal B}\xspace}}

\def\Btopilnu   {\ensuremath{B \to \pi \ell \nu}\xspace}
\def\Btoetalnu  {\ensuremath{B^+ \to \eta \ell^+ \nu}\xspace}
\def\Btoomegalnu  {\ensuremath{B^+ \to \omega \ell^+ \nu}\xspace}
\def\BtoXulnu  {\ensuremath{B \to X_{u} \ell \nu}\xspace}

\def\mes        {\mbox{$m_{\rm ES}$}\xspace}

\def\DeltaE     {\mbox{$\Delta E$}\xspace}

\newcommand{\tev}{\ensuremath{\mathrm{\,Te\kern -0.1em V}}\xspace}
\newcommand{\gev}{\ensuremath{\mathrm{\,Ge\kern -0.1em V}}\xspace}
\newcommand{\mev}{\ensuremath{\mathrm{\,Me\kern -0.1em V}}\xspace}
\newcommand{\kev}{\ensuremath{\mathrm{\,ke\kern -0.1em V}}\xspace}
\newcommand{\ev}{\ensuremath{\mathrm{\,e\kern -0.1em V}}\xspace}
\newcommand{\gevc}{\ensuremath{{\mathrm{\,Ge\kern -0.1em V\!/}c}}\xspace}
\newcommand{\mevc}{\ensuremath{{\mathrm{\,Me\kern -0.1em V\!/}c}}\xspace}
\newcommand{\gevcc}{\ensuremath{{\mathrm{\,Ge\kern -0.1em V\!/}c^2}}\xspace}
\newcommand{\mevcc}{\ensuremath{{\mathrm{\,Me\kern -0.1em V\!/}c^2}}\xspace}

\def\invfb   {\ensuremath{\mbox{\,fb}^{-1}}\xspace}

\def\mus  {\ensuremath{\rm \,\mus}\xspace}

\def\mus        {\ensuremath{\,\mu{\rm s}}\xspace}

\def\ra                 {\ensuremath{\rightarrow}\xspace}
\def\to                 {\ensuremath{\rightarrow}\xspace}

\def\pep2{PEP-II}

\def\gsim{{~\raise.15em\hbox{$>$}\kern-.85em
          \lower.35em\hbox{$\sim$}~}\xspace}
\def\lsim{{~\raise.15em\hbox{$<$}\kern-.85em
          \lower.35em\hbox{$\sim$}~}\xspace}

\xspace

\def\geantfour  {\mbox{\tt GEANT4}\xspace}

\def\jetset74   {\mbox{\tt Jetset \hspace{-0.5em}7.\hspace{-0.2em}4}\xspace}

\newcommand{\etal}{\textit{et al.\@}\xspace}

\def\Xu{\ensuremath{X_u}\xspace}

\def\BXulnu{\ensuremath{B \rightarrow X_u\ell\nu}\xspace}

\def\BtoXclnu{\ensuremath{B \rightarrow X_c\ell\nu}\xspace}

\def\BRBXulnu{\ensuremath{{\cal B}(B \rightarrow X_u\ell\nu)}\xspace}
\def\BRBXclnu{\ensuremath{{\cal B}(B \rightarrow X_c\ell\nu)}\xspace}

\def\BRBpetalnu{\ensuremath{{\cal B}(B^{+} \rightarrow \eta\ell^+\nu})\xspace}

\def\BRBetalnu{\ensuremath{{\cal B}(B^{+} \rightarrow \eta \ell^+\nu})\xspace}
\def\BRBomegalnu{\ensuremath{{\cal B}(B^{+} \rightarrow \omega\ell^+\nu})\xspace}

\def\etagg{\ensuremath{\eta \rightarrow \gamma\gamma}\xspace}
\def\etapipipi{\ensuremath{\eta \rightarrow \pi^+\pi^-\pi^0}\xspace}

\def\DeltaE{\ensuremath{\Delta E}\xspace}

\def\cosBYDef{\ensuremath{\cos \theta_{BY} = \left(2 E^*_B
      E^*_Y-m_B^2-m_Y^2\right)/\left(2|\vec p^{\,*}_B|
      |\vec p^{\,*}_Y|\right)}\xspace}
\def\mESDef{\ensuremath{m_{\rm ES} = \sqrt{(s/2+\vec{p}_B \cdot \vec{p}_{\rm beam})^2/E_{\rm beam}^2- \vec{p}_B^{\,2}}}\xspace}
\def\DeltaEDef{\ensuremath{\Delta E = (P_B \cdot P_{\rm beam} - s/2) / \sqrt{s}}\xspace}

\def\ThetamissCut{\ensuremath{0.3 < \theta_{\rm miss} < 2.2}~\rm~rad\xspace}

\def\plepCut{\ensuremath{|\vec{p}^{\,*}_{\rm \ell}| > 1.6~(1.0)~\gev}\xspace}

\def\mmissEmissCut{\ensuremath{|m_{\rm miss}^2 / (2 E_{\rm miss}) | < 2.5~\gev }\xspace}
\def\cosBYCut{\ensuremath{-1.2 < \cos \theta_{\rm BY} < 1.1}\xspace}

\def\FitRegion{\ensuremath{|\Delta E| < 0.95~\gev \mbox{ and } m_{\rm ES} > 5.095~\gev}\xspace}

\def\SignalBandDeltaE{\ensuremath{-0.2 < \Delta E < 0.4~\gev}\xspace}
\def\SignalBandmES{\ensuremath{m_{\rm ES} > 5.255~\gev}\xspace}

\begin{document}

\preprint{\babar-PUB-\BABARPubYear/\BABARPubNumber} 
\preprint{SLAC-PUB-\SLACPubNumber} 
\preprint{arXiv: \arxiv} 

\begin{flushleft}
\babar-PUB-\BABARPubYear/\BABARPubNumber\\
SLAC-PUB-\SLACPubNumber\\
arXiv: \arxiv\\
\end{flushleft}

\title{\large\bf 
Measurement of the \boldmath \Btoomegalnu and \Btoetalnu Branching Fractions}

%
\author{B.~Aubert}
\author{M.~Bona}
\author{Y.~Karyotakis}
\author{J.~P.~Lees}
\author{V.~Poireau}
\author{E.~Prencipe}
\author{X.~Prudent}
\author{V.~Tisserand}
\affiliation{Laboratoire de Physique des Particules, IN2P3/CNRS et Universit\'e de Savoie, F-74941 Annecy-Le-Vieux, France }
\author{J.~Garra~Tico}
\author{E.~Grauges}
\affiliation{Universitat de Barcelona, Facultat de Fisica, Departament ECM, E-08028 Barcelona, Spain }
% ...
\collaboration{The \babar\ Collaboration}
\noaffiliation

\date{\today}

\begin{abstract}
We present a study of the charmless semileptonic $B$-meson decays
\Btoomegalnu and \Btoetalnu. The analysis is 
based on $383$ million $B\overline B$ pairs recorded at 
the $\Upsilon(4S)$ resonance with the \babar\ detector.
The $\omega$ mesons are reconstructed in the channel $\omega \rightarrow \pi^+\pi^-\pi^0$ and
the $\eta$ mesons in the channels $\eta \rightarrow \pi^+\pi^-\pi^0$
and $\eta \rightarrow \gamma \gamma$. We measure the branching fractions 
\BRBomegalnu = $(1.14 \pm 0.16_{\mathrm stat} \pm 0.08_{\mathrm syst})\times10^{-4}$ 
and 
\BRBetalnu = $(0.31 \pm 0.06_{\mathrm stat} \pm 0.08_{\mathrm syst}  ) \times 10^{-4}$.
\end{abstract}

 \pacs{13.20.He,                
       14.40.Nd}                

\maketitle

Measurements of branching fractions of charmless semileptonic $B$ decays can be used to
determine the magnitude of the Cabibbo-Kobayashi-Maskawa matrix~\cite{CKM} element
$V_{ub}$ and thus provide an important constraint on the Unitarity Triangle.
Studies of exclusive decays allow for more stringent kinematic constraints
and better background suppression than inclusive measurements.
However, the predictions for exclusive decay rates depend on calculations of 
hadronic form factors and 
are thus affected by theoretical uncertainties different from those
involved in inclusive decays. 
The description of semileptonic decays 
requires one or three form factors for final states with a pseudoscalar 
or a vector meson, respectively, if
lepton masses are neglected.
Currently, the most precise determination of $|V_{ub}|$
with exclusive decays, both experimentally and theoretically,
comes from a measurement of \Btopilnu decays~\cite{babarpilnu}. 
It is important to study other semileptonic final states with a
pseudoscalar or a vector meson 
to perform further tests of theoretical calculations and to improve the 
knowledge of the composition of charmless semileptonic decays.

In this paper, we present measurements of the branching fractions 
\BRBomegalnu and \BRBpetalnu, where $\ell = e, \mu$
and charge-conjugate modes are included implicitly.
These decays have previously been studied by 
the CLEO~\cite{cleo} and \babar~\cite{babartagged, babarsltag} 
collaborations (\Btoetalnu) 
and by the Belle~\cite{belle} collaboration (\Btoomegalnu).
The $\omega$ meson is reconstructed  
in its decay to three pions 
(${\cal B} (\omega \rightarrow \pi^+\pi^-\pi^0)= (89.2\pm 0.7)\%$
\cite{PDG}), while for the $\eta$ meson the decays to three pions
and to two photons
($ {\cal B} (\eta \rightarrow \pi^+\pi^-\pi^0) =(22.68\pm 0.35)\%$,  
 $ {\cal B} (\eta \rightarrow \gamma \gamma)  = (39.39 \pm 0.24)\% $ 
\cite{PDG}) are used. 
In contrast to earlier \Btoetalnu analyses from 
\babar~\cite{babartagged, babarsltag}, 
the second $B$ meson in the event is not reconstructed; this 
yields a much larger candidate sample. 

The results presented here are based on a data sample of 
$383$ million \BB\ pairs recorded with the \babar\ detector~\cite{babar}
at the PEP-II asymmetric-energy \epem storage rings at the Stanford Linear Accelerator Center (SLAC).  
The data correspond to an integrated luminosity of $347$~\invfb
collected at the \FourS resonance. In addition, $35$~\invfb of data collected about $40$
\mev\ below the resonance (off-resonance) are used for background studies.
Simulated \BB\ events are used to estimate signal efficiencies and 
shapes of signal and background distributions. Charmless semileptonic
decays are simulated as a mixture of 
three-body decays \BXulnu ($\Xu = \pi, \eta, \eta', \rho, \omega$)
and have been reweighted according to the latest form-factor calculations 
from light-cone sum rules~\cite{lcsr:pi, lcsr:rho, lcsr:eta}. 
Decays to non-resonant hadronic states \Xu with masses $m_{\Xu} > 2m_\pi$ 
are simulated using the differential decay rate given in Ref.~\cite{DFN},
which produces a smooth $m_{\Xu}$ spectrum.
The \geantfour package~\cite{geant} is used to model
the \babar\ detector response.

The reconstruction of the signal decays 
\Btoomegalnu  and \Btoetalnu 
requires the identification of a charged lepton ($e$ or $\mu$) and the
reconstruction of an $\omega$ or $\eta$ meson.
The two dominant sources of background are semileptonic decays with a charm meson
in the final state, \BtoXclnu ($X_c = D, D^{*}, D^{**}, D^{(*)}\pi$),
and $e^+e^-\to q\bar{q}$~$(q=u,d,s,c)$ continuum events.
Other backgrounds include charmless semileptonic decays that are not analyzed
as signal and
\BB events with lepton candidates from secondary decays or from misidentification of
hadrons as leptons.
The center-of-mass momentum of the lepton is restricted to \plepCut \cite{Vstar} 
for the $\omega$ ($\eta$) final state.
This requirement 
significantly reduces those background events that have hadrons misidentified as leptons
 and rejects a large fraction of leptons from secondary decays or photon 
conversions.
For the reconstruction of the $\omega$ or $\eta$ meson, charged
(neutral) pions are required to have a momentum in the laboratory frame above 
$200$~($400$)~\mev to reduce combinatorial background. 
Neutral pion candidates are formed from two photons with energies above
$100$~\mev and an invariant mass, $m_{\gamma\gamma}$, 
in the range $100 < m_{\gamma\gamma} < 160 \mev $. 
A three-pion system is accepted as an $\omega$ ($\eta$) candidate
if its invariant mass, $m_{3\pi}$, is in the range $760 <
m_{3\pi}< 806\mev$ for $\omega$ candidates 
and  $540 < m_{3\pi}< 555\mev$ for $\eta$ candidates. 
The $\eta$ meson is also reconstructed via its decay into two photons,
each with an energy above $50$~\mev, with a two-photon
invariant mass in the range $520 < m_{\gamma\gamma}< 570\mev$. 
To reduce the combinatorial background, two-photon
combinations are rejected as possible $\eta$
candidates if either of the photons can be combined with any other 
photon of the event to form a system with an invariant mass close to
the $\pi^0$ mass, $110 < m_{\gamma\gamma}< 160\mev$.

Event-shape variables that are sensitive to the topological differences
between jet-like continuum events and more spherical \BB events
are used to suppress backgrounds from $e^+e^- \ra \qqbar$ and other QED processes. 
The normalized second Fox-Wolfram moment $R_2$~\cite{R2} is required to be less
than $0.5$ and a loose requirement on the second Legendre moment $L_2$~\cite{L2} 
of $L_2 < 3.0 \gev$ is imposed. In addition, the event must contain at least 
four charged tracks.

The charged lepton is combined with an $\omega$($\eta$) candidate
to form a so-called $Y$ pseudo-particle candidate, whose four-momentum is
defined as the sum of the corresponding lepton and hadron four-momenta.
All charged tracks belonging to the $Y$ are fit to a common vertex.
This vertex fit must yield a $\chi^2$ probability of at least 0.1\%.
Multiple $Y$ candidates per event are possible and all candidates are retained. 
The $Y$ multiplicity  is
well described by the Monte-Carlo simulation.
About 96\% (98\%) of 
simulated \Btoomegalnu (\Btoetalnu) signal events and more than 
90\% of all selected data events, which include both signal and background, 
contain only one $Y$ candidate. 

The momentum of the candidate
neutrino is calculated from the difference between the momenta of the colliding-beam particles
and the vector sum of the momenta of all detected particles in the
event. 
The energy of the candidate neutrino is obtained as
the magnitude of its momentum, since this is less susceptible to bias
from lost particles or additional tracks than the missing energy, $E_{\rm miss}$, of the event.
The magnitude of the missing-momentum vector must be at least $500~\mev$.
The effect of losses due to detector acceptance 
on the reconstruction of the neutrino candidate is reduced by requiring
the missing-momentum vector in the laboratory frame to point into
the polar-angle range \ThetamissCut. 
If the missing energy and momentum in the event come from 
a single undetected neutrino and the rest of the event is correctly
reconstructed, the missing mass, $m_{\rm miss}$, measured from the
whole event should be compatible with zero.  
Because the missing-mass resolution varies
linearly with the missing energy, only events with \mmissEmissCut
are selected.

If the $Y$ candidate originates from a signal decay that has
been correctly reconstructed, the cosine of the angle between the $B$ meson
and the $Y$ candidate can be calculated as \cosBYDef. 
Here $m_B, E^*_B, \vec p^{\,*}_B, m_Y,  E^*_Y, \vec p^{\,*}_Y$ 
refer to the masses, energies, and momenta of the $B$ meson and the $Y$ candidate, respectively.
In the calculation of $\cos \theta_{BY}$, 
the $B$-meson energy $E^*_B$ and momentum $\vec p^{\,*}_B$ are not
measured event by event. Instead, $E^*_B = \sqrt{s}/2$ is given
by the center-of-mass energy of the colliding beam particles, $\sqrt{s}$, 
and the magnitude of the
$B$ momentum is calculated as $|\vec p^{\,*}_B| = \sqrt{E^{*2}_B-m^2_B}$.
Signal candidates are required to satisfy \cosBYCut. 
This requirement was kept loose to account for the limited detector resolution and photon energy losses.

To reduce backgrounds without significant loss of signal, 
the momenta of the lepton, $\vec{p}_\ell^*$,
and of the hadron, $\vec{p}_{\omega,\eta}^*$, that make up a $Y$ candidate 
are restricted. 
For \Btoomegalnu , the momenta are required to satisfy
$|\vec{p}_{\omega}^*| > 1.3~\gev$ or  $|\vec{p}_\ell^*| > 2.0~\gev$
or $|\vec{p}_{\omega}^*|  +  |\vec{p}_\ell^*| > 2.65~\gev$. 
In the case of \Btoetalnu , the conditions 
$|\vec{p}_{\eta}^*| > 1.3~\gev$ or $|\vec{p}_\ell^*| > 2.1~\gev$
or $|\vec{p}_{\eta}^*|  +  |\vec{p}_\ell^*| > 2.8~\gev$ have to be
fulfilled. 

The kinematic consistency of the reconstructed $Y\nu$ system with a signal $B$ decay
is verified using the two variables \DeltaEDef and \mESDef, where
$P_{\rm beam}=(E_{\rm beam},\vec{p}_{\rm beam})$ is 
the four-momentum of the colliding beam particles and
$P_B= (E_B, \vec{p}_B)$ is the $B$-meson four-momentum 
computed as the sum of the four-momenta of the $Y$ and the $\nu$ candidates.
These variables are later used to extract the signal yields
in a fit to the two-dimensional \DeltaE~vs.~\mes distribution.
Only candidates that fulfill the loose requirement \FitRegion\ (fit region) are retained. 
 
At this stage of the selection, 
the signal-to-background ratio, $S/B$, 
where $S$ and $B$ denote the expected signal and background yields, respectively,
is small.
It amounts to 1.5\% for \Btoomegalnu,
and 1.8\% (1.0\%)  for \Btoetalnu  with 
$\eta\rightarrow \pi^+\pi^-\pi^0 (\gamma\gamma)$. 
The signal efficiencies for the sum of decays with electrons and muons,
estimated from simulation, are
2.8\% for \Btoomegalnu and  4.0\% (9.4\%) for \Btoetalnu 
with $\eta\rightarrow \pi^+\pi^-\pi^0 (\gamma\gamma)$.

For further discrimination between the signal and the background, 
a multivariate selection based on neural networks~\cite{TMVA} is used. 
For each of the three signal channels under study,  
neural networks with two hidden layers (four and two
neurons, respectively) are applied consecutively to separate the signal
from the two main backgrounds. A first neural network discriminates
the signal against \qqbar continuum events; a second network is used to 
further distinguish the signal from the \BtoXclnu background. 
The neural-network decision is based on the following input variables:
$m_{miss}^{2}/(2 E_{miss})$, $\theta_{miss}$, 
$\cos\theta_{BY}$, $R_2$, $L_2$,
$\cos\Delta\theta_{thrust}$,
the cosine of the polar-angle difference between the thrust axes
of the $Y$ candidate and of the rest of the event,
and $\cos\theta_{W\ell}$,  the cosine of the lepton ``helicity angle''
measured in the rest frame of the virtual $W$ 
(calculated using the lepton and neutrino candidates) relative to the 
$W$ direction in the $B$ rest frame.
For the three-pion final states, the Dalitz 
amplitude, the magnitude of the vector product of the  $\pi^+$ momentum 
and the $\pi^-$ momentum in the $\omega/\eta$
rest frame, normalized to its maximum value, serves as an additional
input variable to separate $\omega \rightarrow \pi^+\pi^-\pi^0$ 
and $\eta \rightarrow \pi^+\pi^-\pi^0$ decays from combinatorial background.

The training of the neural networks is done using the corresponding simulated 
signal and background samples for each of
the three signal channels separately. Independent simulated event samples
are used to validate the training. 
Based on Monte-Carlo simulation, the selection criterion
for each of the output discriminants is chosen to
maximize the quantity
$S/\sqrt{S+B}$.
The signal efficiencies and the $S/B$ ratios 
after the neural-network selection are given in 
Table~\ref{Tab:Selection} for the fit region
and for the signal region 
delimited by $-0.2 < \DeltaE < 0.4 \gev$ and $ m_{ES} > 5.255 \gev$.

\begin{table}
\caption{Signal efficiencies, $\epsilon_{signal}$, 
and signal-to-background ratios, $S/B$, after the neural-network selection.}
\label{Tab:Selection}
\begin{tabular}{|l|c|c|c|} \hline
             & \multicolumn{2}{c|}{Fit region} & \multicolumn{1}{c|}{Signal region}\\
             &$\epsilon_{\mathrm signal}(\%)$&$S/B$ & $S/B$\\
\hline
\Btoomegalnu             & 1.00 & 0.15 & 0.46 \\
\Btoetalnu, \etapipipi   & 1.62 & 0.10 & 0.35 \\
\Btoetalnu, \etagg       & 4.90 & 0.04 & 0.15 \\ \hline
\end{tabular}
\end{table}

For the determination of the signal branching fractions,  
the \DeltaE~vs.~\mes distributions of the
simulated signal and backgrounds are fit to the data
distribution for the three signal channels independently.
The fits are based on an extended binned maximum-likelihood
method~\cite{bbfit} and take statistical fluctuations of both 
the data and the Monte-Carlo samples into account.
The binning of the \DeltaE~vs.~\mes distributions used in the fits contains 
a total of $50$ bins with smaller sizes in the signal region 
to resolve the signal shape and larger sizes
in the part of the fit region outside of the signal region 
to determine the background normalizations from data.
The shapes of the signal and background distributions are taken from simulation.
The fits determine the relative fractions of the signal 
and some of the background samples in the data. 

The free parameters of the fits are the normalizations of the 
signal and the $B\rightarrow X_c\ell\nu$ background, 
and for the \Btoomegalnu channel also the overall normalization
of the continuum background. 
The \BtoXclnu normalization is left free to account 
for a slight discrepancy between the \BtoXclnu yields in   
data and Monte-Carlo simulation. 
The relative contributions of events with electrons or muons 
to the continuum background have been determined
using off-resonance data. 
Compared to the \Btoomegalnu channel, the \Btoetalnu channels
suffer from a larger 
continuum background and the fit shows larger correlations between signal and this 
background component. The normalization of the continuum background in the \Btoetalnu channel 
is therefore taken from off-resonance data and is not varied in the fit.
All other background distributions, mainly other \BXulnu\ decays 
and $B$ decays with secondary or misidentified leptons, 
are fixed to their Monte-Carlo predictions. 
The fit procedure has been validated with simulated signal
and background samples. 

The resulting signal yields and branching fractions for the three signal 
channels are presented in Table~\ref{Tab:Fitresult}.
The scale factors between the $B\rightarrow X_c\ell\nu$ background yields
predicted by the simulation and the values determined by the fits are
$1.06 \pm 0.07$ for \Btoomegalnu and $0.96 \pm 0.07$ ($1.12\pm0.03$)
for \Btoetalnu with \etapipipi (\etagg). 
The correlations between the signal and the $B\rightarrow X_c\ell\nu$ 
parameters determined by the fits 
are $0.08$ for \Btoomegalnu and $-0.60$ ($-0.46$) 
for \Btoetalnu with \etapipipi (\etagg).
The correlation between the signal and the continuum parameters for
the \Btoomegalnu channel is $-0.55$ and the continuum background normalization is
adjusted by a factor of $0.89 \pm 0.12$ with respect to the
normalization obtained from the off-resonance data sample.
The goodness of fit is evaluated using a $\chi^2$-based comparison of the
fitted \DeltaE~vs.~\mes distributions of the simulated and data samples and is
shown in Table~\ref{Tab:Fitresult}.
In addition, the combined branching fraction for the two \Btoetalnu
channels has been obtained from a fit to the sum of the
\DeltaE~vs.~\mes distributions for \etapipipi and \etagg.
 
\begin{figure*}[htb]
\begin{center}
\noindent
\hskip -1.0 cm
\epsfig{file=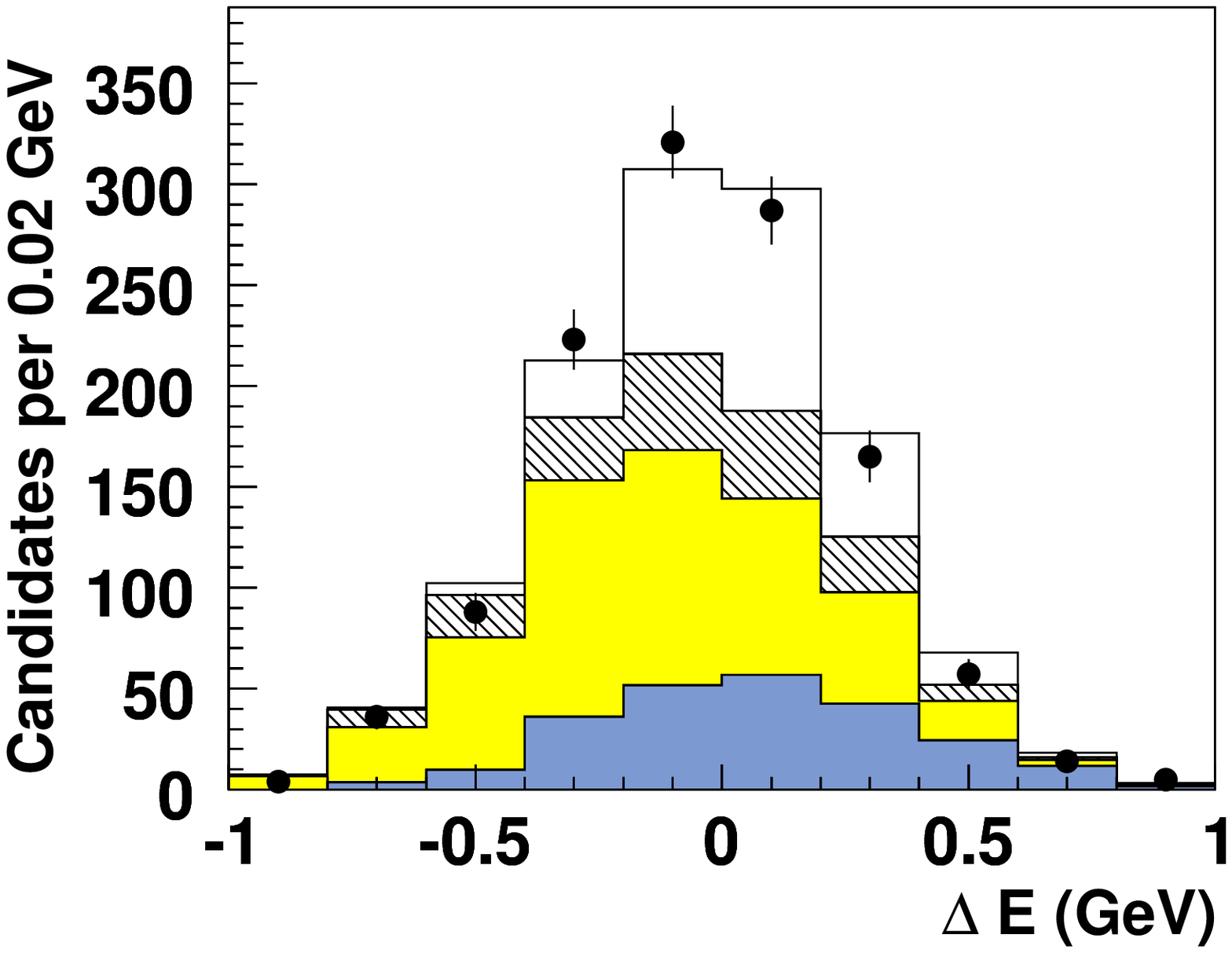, width=5.4cm}
\epsfig{file=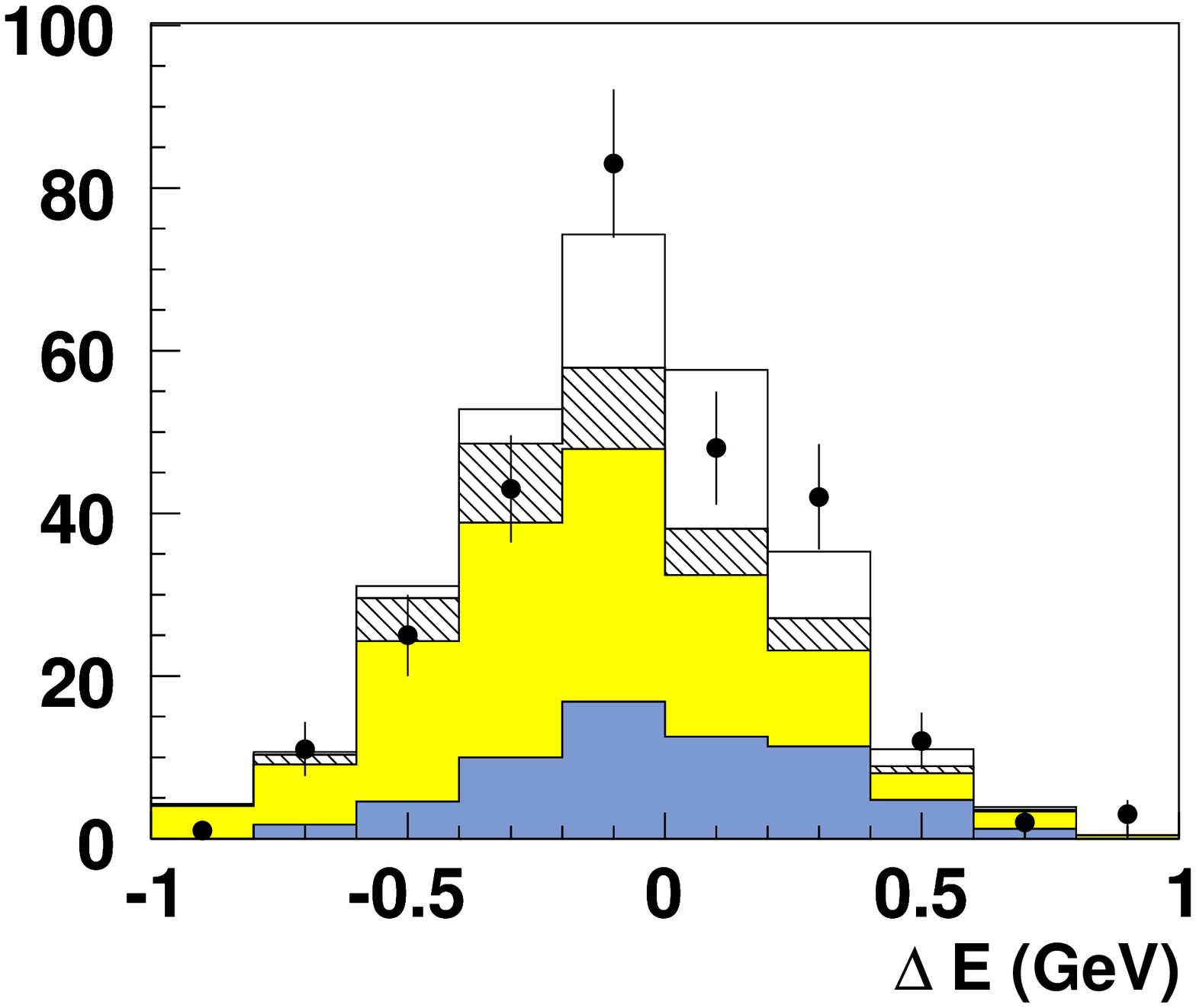, width=4.35cm}
\epsfig{file=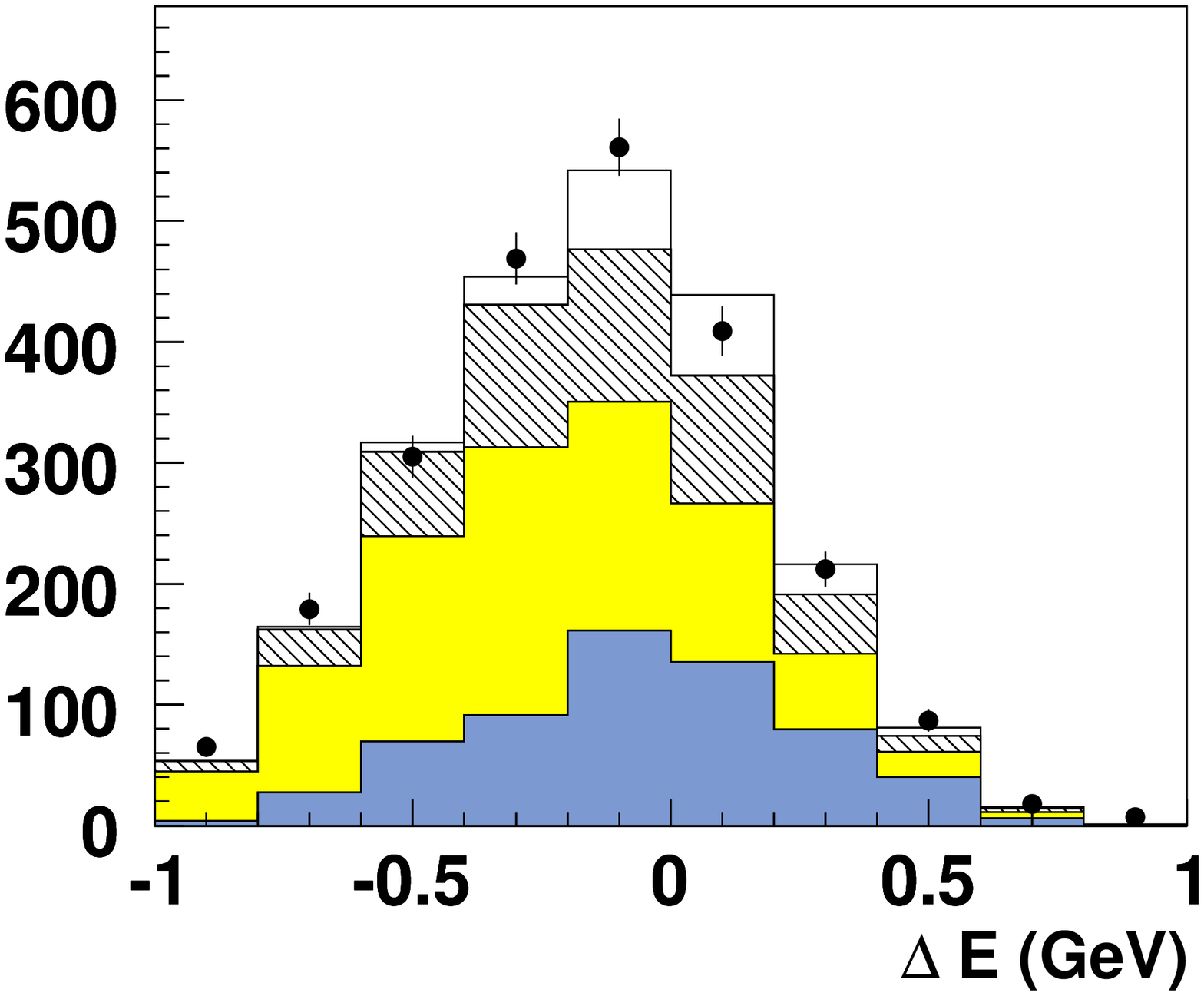, width=4.35cm}
\epsfig{file=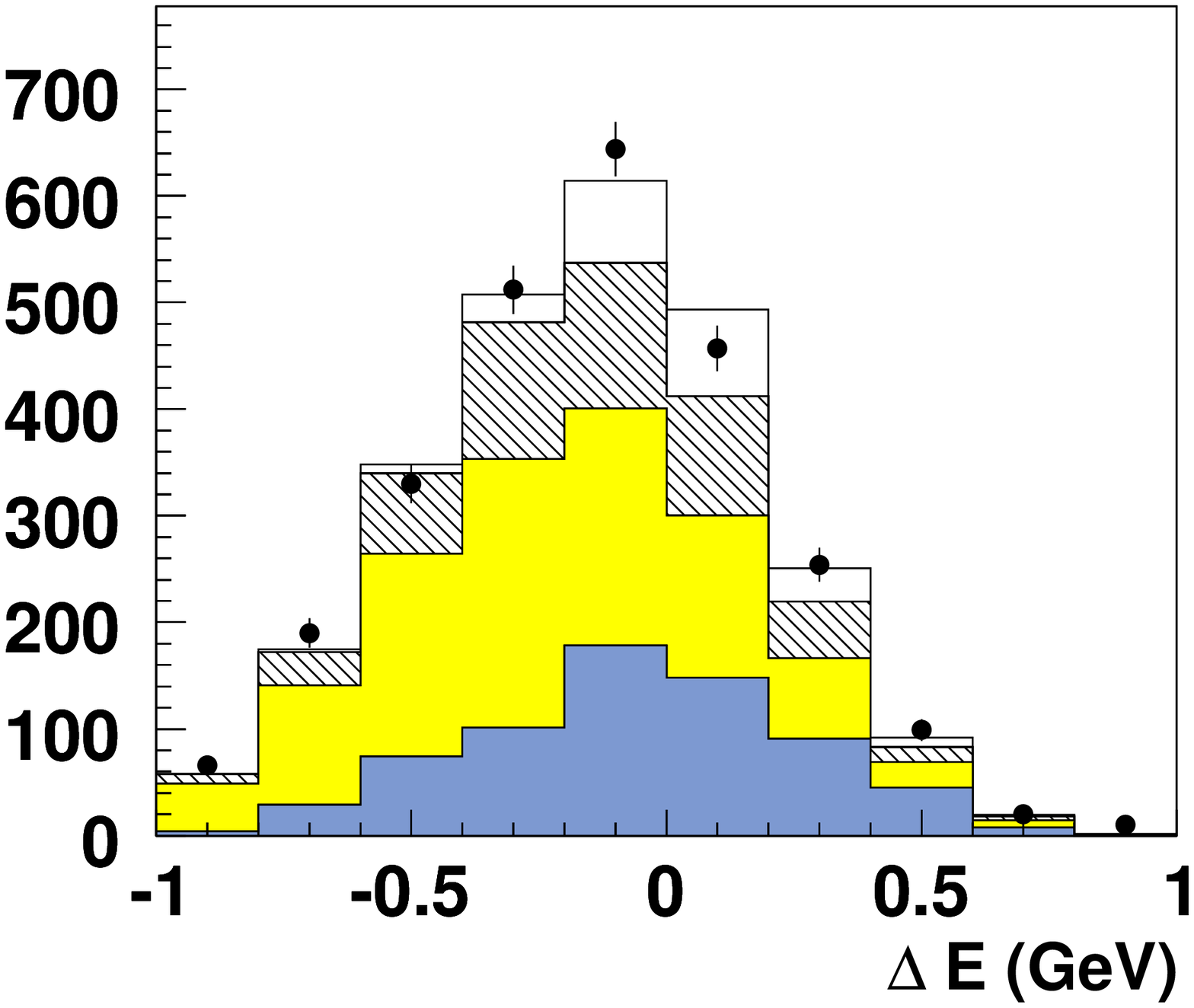, width=4.35cm}\\ \vskip 0.1cm
\hskip -1.0 cm
\epsfig{file=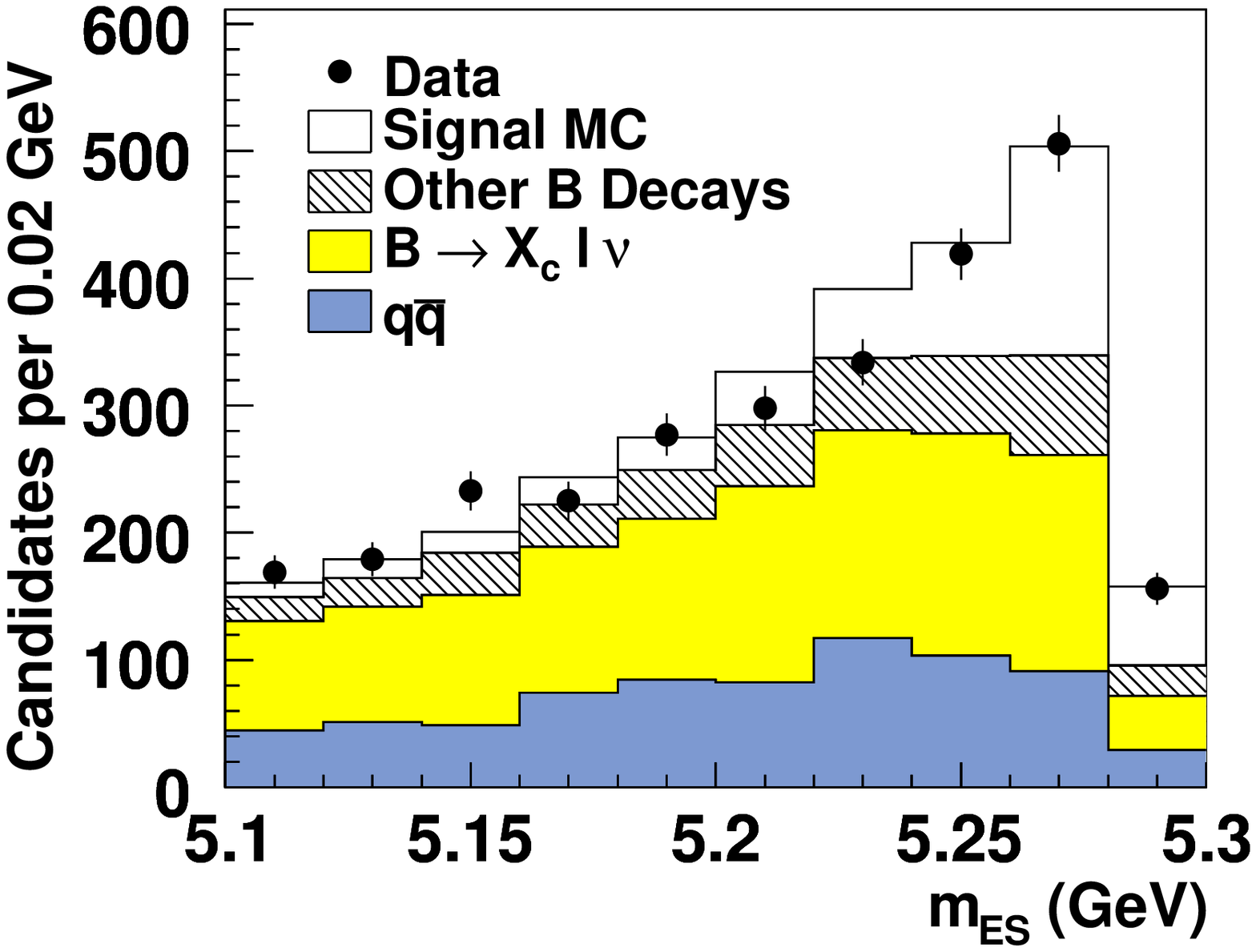, width=5.4cm}
\epsfig{file=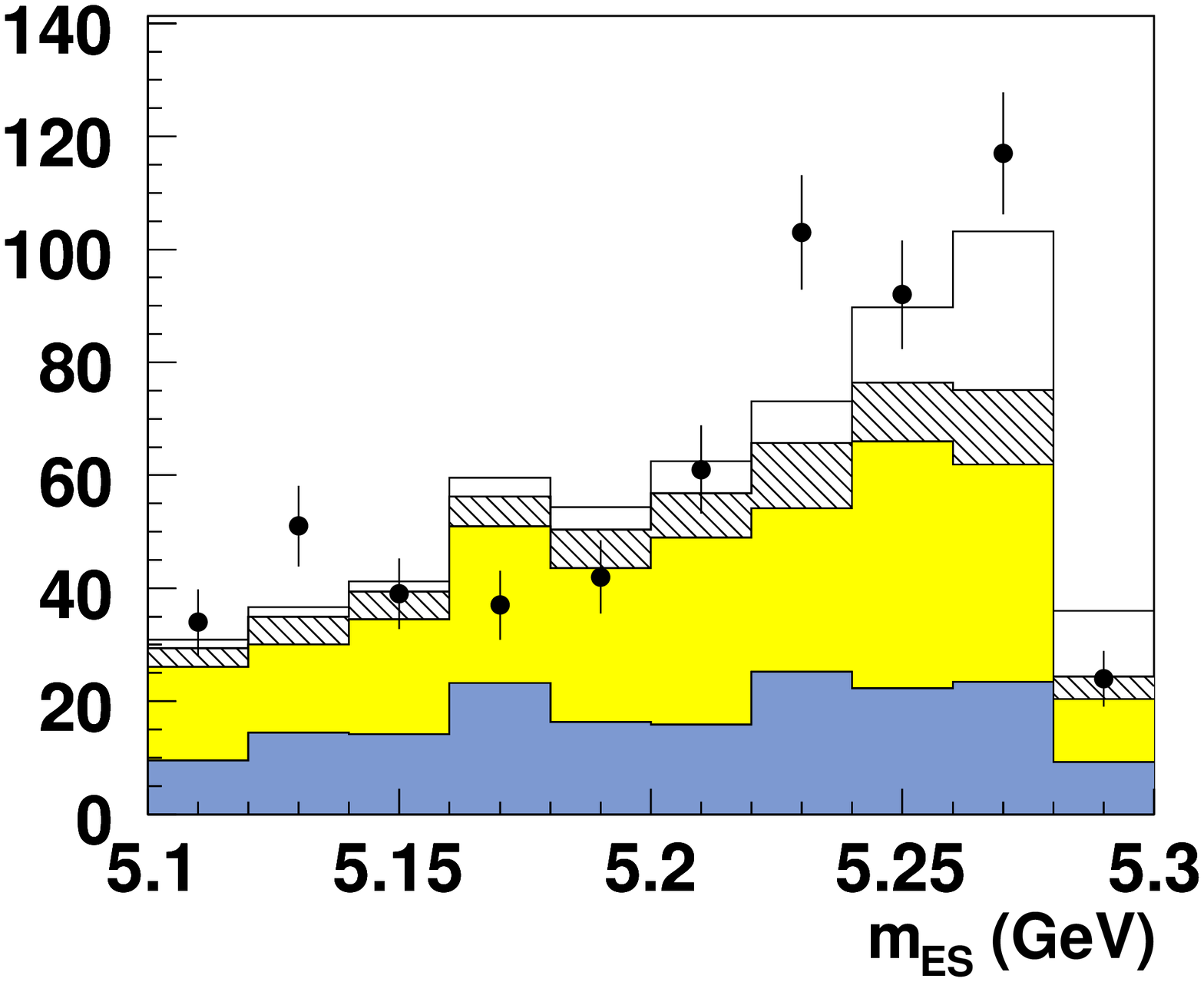, width=4.35cm}
\epsfig{file=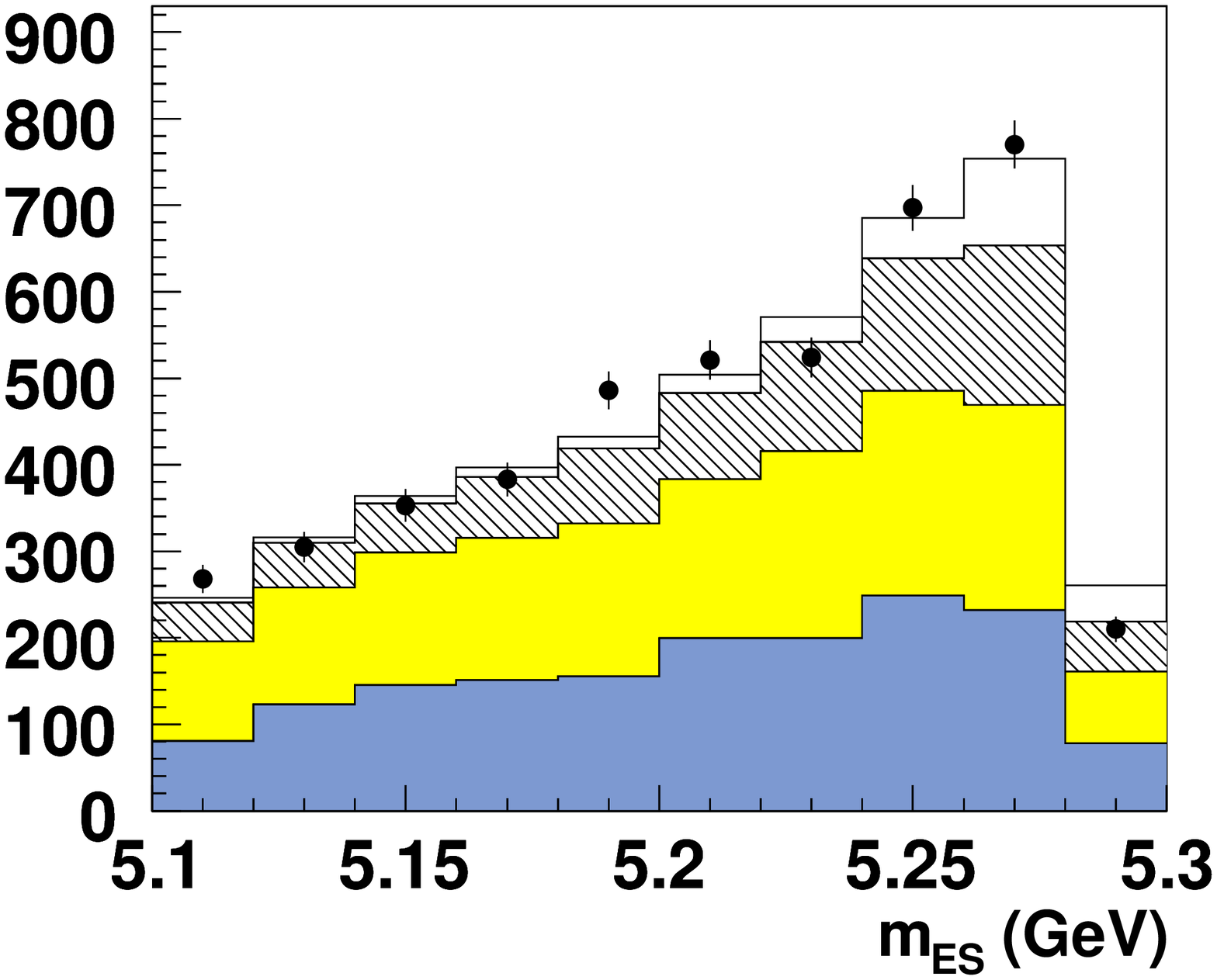, width=4.35cm}
\epsfig{file=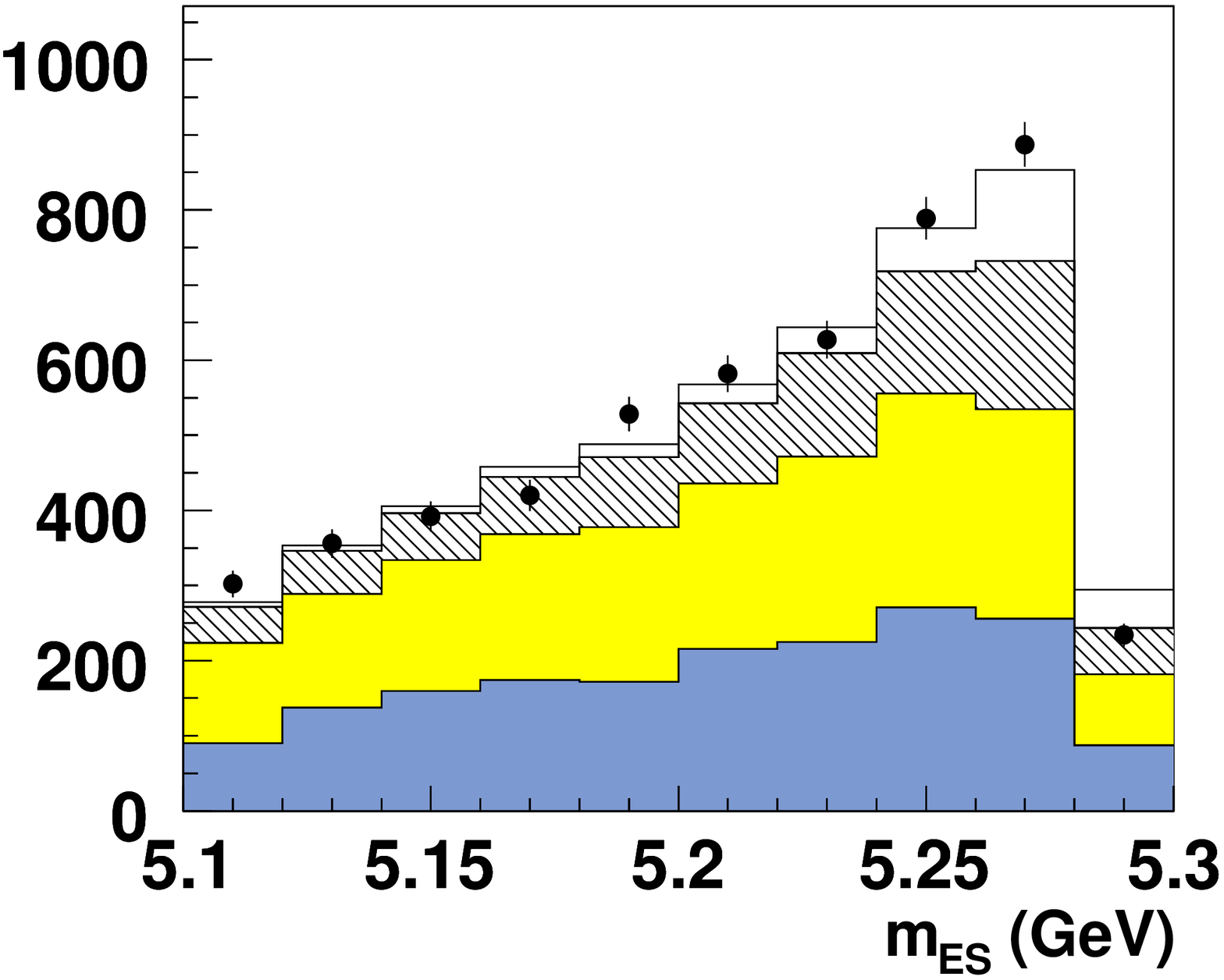, width=4.35cm}
\caption{
(color online) Projected $\DeltaE$ distributions for \SignalBandmES (top), 
and $m_{ES}$ distributions for \SignalBandDeltaE (bottom).
 From left to right: \Btoomegalnu channel,
\Btoetalnu channel with $\eta\rightarrow\pi^+\pi^-\pi^0$, 
\Btoetalnu channel with $\eta\rightarrow\gamma\gamma$,
and combined \Btoetalnu channel.
The error bars represent the statistical uncertainties on the data.
The histograms show simulated distributions for signal (white), \BtoXclnu decays
(light shaded/yellow), \qqbar-continuum (dark shaded/blue) and all other backgrounds (hatched)
and have been summed up.
The distributions of the simulated signal and \BtoXclnu background 
(and the $q\bar{q}$ background for \Btoomegalnu) have been scaled to the results 
of the fits. 
}
\label{Fig:SignalFit} 
\end{center}
\end{figure*}

Figure~\ref{Fig:SignalFit} shows the projections of the fitted
distributions on $\DeltaE$ and $\mes$ for the three signal channels and the
combined \Btoetalnu channel.
For a better visibility of the signal, 
the $\Delta E$ projections are shown for $m_{ES} > 5.255 \gev$ and
the $m_{ES}$ projections are shown for  $-0.2 < \DeltaE < 0.4 \gev$.

\begin{table}
 \renewcommand{\arraystretch}{1.0}
\caption{Signal yields and corresponding branching fractions as determined by the fits 
for the three signal channels and the combined \Btoetalnu channel.
The last row shows the $\chi^2$ per degree of freedom.}
\label{Tab:Fitresult}
\begin{tabular}{|l|c|ccc|}
\hline 
   & \multicolumn{1}{c|}{\Btoomegalnu} & \multicolumn{3}{c|}{\Btoetalnu  } \\
   &           &  \etapipipi \  & \etagg \ & combined \\
\hline
$N_{signal}$               & $802\pm113$       & $127\pm42$                  
                           & $459\pm98$        & $554\pm105$\\
\BR  ($10^{-5}$)           & $ 11.4\pm 1.6   $ & $4.36 \pm 1.43 $ 
                           & $ 3.01\pm 0.64  $ & $3.11\pm 0.59 $\\
$\chi^2/d.o.f.$            & $36.0/47$ & $59.9/48$ & $43.2/48$ & $49.7/48$\\
\hline
\end{tabular}
\end{table}

The systematic errors on the measured branching fractions are listed in
Table~\ref{Tab:Systematics}. They are estimated by varying 
the detection efficiencies or the parameters that impact the
modeling of the signal and the background processes
within their uncertainties.
The complete analysis is then repeated and the differences in the 
resulting branching fractions are taken as the systematic errors.
The total systematic error is obtained by adding in quadrature 
all listed contributions.

Uncertainties due to the reconstruction of charged particles and photons 
are evaluated by varying their reconstruction efficiencies 
and the energy depositions of photons in the simulation. 
The neutrino reconstruction is affected by background with 
long-lived $K^0_L$, which often escape detection and
contribute to the measured missing momentum of the event. 
The uncertainty arising from $K_L^0$ production and interactions
is estimated by varying their production rate as well as their 
detection efficiency and energy deposition in the simulation. 
For lepton identification, relative uncertainties of $1.4\%$
and  $3\%$ are used for electrons and muons, respectively. 
A $3\%$ uncertainty is assigned to the
$\pi^0/\eta\rightarrow\gamma\gamma$ reconstruction efficiency.

The uncertainty due to the \BtoXclnu background is evaluated by varying
the $ B\rightarrow D/D^*/D^{**} \ell\nu $ branching fractions~\cite{PDG} 
and the $B\rightarrow D^*$ form factors~\cite{DstarFF}. 
Prior to the neural-network selection, the background 
level is high and discrepancies between data and Monte-Carlo
distributions are observed at roughly the 10\% level 
for some of the neural-network input variables. 
To estimate the effect of these discrepancies on the measured branching 
fractions, the dominant background component (\BtoXclnu) is reweighted to match the data. 
The weights are determined from a \BtoXclnu-enhanced sample which is  
obtained by selecting only events that are otherwise rejected by the 
\BtoXclnu neural-network selection and keeping all other selection
criteria unchanged.

For the \BtoXulnu background, the branching fractions of the 
exclusive decays that are not analyzed as signal are varied
within their uncertainties~\cite{HFAG}.
The non-resonant part is varied
within the range allowed by the uncertainty of the total \BtoXulnu
branching fraction~\cite{HFAG}.
The uncertainty due to the normalization of the continuum background 
has been determined with off-resonance data for events with electrons or muons
separately. 
Since the overall normalization of the continuum background is 
adjusted for \Btoomegalnu in the fit, the resulting error in this channel 
is smaller than for the \Btoetalnu channels.
For the normalization of 
secondary-lepton background, an uncertainty of
6-8\%, depending on the signal channel, 
has been estimated from a detailed study of the composition 
of this background.

Uncertainties in the modeling of signal decays due to the imperfect
knowledge of the form factors affect
the shapes of kinematic spectra and thus the acceptances of signal decays. 
The errors on the measured branching fractions
are estimated by varying the parameters of the form-factor calculations
withing their uncertainties~\cite{lcsr:rho, lcsr:eta}.

The branching fractions of
the decays $\omega/\eta \rightarrow \pi^+\pi^-\pi^0$
and $\eta \rightarrow \gamma\gamma$ are also varied within their
uncertainties~\cite{PDG}. 
The uncertainty on the number of produced $B$ mesons 
is $1.1\%$~\cite{Bcounting}. 
The uncertainty on the ratio of the $\Upsilon(4S)\to B^+B^-$ 
and $\Upsilon(4S)\to B^0\overline{B^0}$ branching fractions,
$f_{+-}/f_{00} = 1.065 \pm 0.026~$\cite{HFAG}, 
is taken into account.

The total systematic errors on the measured branching fractions are
7.2\% and 25.1\% for the \Btoomegalnu and the combined \Btoetalnu
channels, respectively.

\begin{table*}[htbp]
\begin{center}
\caption{Relative systematic errors of the branching
  fractions \BRBomegalnu and \BRBetalnu. For the \Btoetalnu channel, the
  systematic errors for the three-pion and the two-photon final states as well
  as for the combined result are shown.
  The total error in each column is the sum in
  quadrature of all listed contributions.}
\label{Tab:Systematics}

\renewcommand{\arraystretch}{1.0}
\begin{tabular}{l|c|ccc}
\hline
 Error source & \multicolumn{1}{c|}{$\delta\BRBomegalnu~(\%)$} & \multicolumn{3}{c}
{$\delta\BRBetalnu~(\%)$ } \\
  &            &  \etapipipi \  & \etagg \ & combined \\
\hline
Tracking efficiency                  & 1.9 & 4.9 & 4.2 & 4.6  \\  
Photon reconstruction                & 2.1 & 1.8 & 9.1 & 8.6  \\ 
$K^0_L$ production and interactions  & 2.6 & 4.8 & 3.1 &  1.9   \\
Lepton identification                & 1.9 & 3.3 & 6.9 &  6.3  \\
$\pi^0/\eta$ identification          & 3.8 & 6.9 & 13.3 &  12.2  \\ 
Neural-net input variables           & 0.6 & 0.8 & 5.9 &   6.1   \\
\hline
$D^{*}$ form factors                 & 0.4 & 1.0 & 0.9 &  1.0 \\
\BRBXclnu                            & 2.1 & 5.5 & 7.6 &  8.0   \\
\BRBXulnu                            & 2.8 & 4.4 & 9.8 &  8.6  \\
Secondary leptons                    & 0.3 & 0.3 & 0.6 & 0.5   \\
Continuum scaling                    & 0.7 & 15.8 & 10.4 &  12.7   \\    
Signal form factor(s)                & 1.8 & 5.9 & 0.3 &  1.3 \\
${\cal B}(\omega/\eta\rightarrow \pi^+\pi^-\pi^0)$, ${\cal B}(\eta\rightarrow \gamma \gamma)$ & 0.8 & 1.5 & 0.6 &  1.2    \\
$N_{B\bar{B}}$                       & 1.1 & 1.1 & 1.1 &  1.1  \\
$f_{+-}/f_{00}$                      & 1.2 & 1.4 & 0.9 &  1.0  \\
\hline
Total systematic error               & 7.2 & 21.1 & 25.2 & 25.1 
\\ \hline
\end{tabular}
\end{center}
\end{table*}

In summary,
we have measured the branching fractions of \Btoomegalnu and
\Btoetalnu decays to be
\begin{eqnarray*}
  \BRBomegalnu &=& (1.14 \pm 0.16 \pm 0.08)\times 10^{-4} , \\
  \BRBetalnu   &=& (0.31 \pm 0.06 \pm 0.08)\times 10^{-4} ,
\end{eqnarray*}
where the errors are statistical (data and simulation) and systematic, respectively. 

The \Btoetalnu and \Btoomegalnu measurements presented here
significantly improve the current knowledge of these decays.
The \Btoetalnu result is compatible with an earlier measurement
by \babar~\cite{babarsltag}
based on events tagged by a semileptonic decay of the second $B$ meson,
$\BRBetalnu = \left(0.64\pm0.20_{stat}\pm0.03_{syst}\right)\times 10^{-4}$.
The two analyses are statistically independent and complement each other.
The analysis presented here is statistically more precise but has larger
systematic uncertainties, as expected for an untagged measurement.
We combine the two \babar\ results and obtain
$\BRBetalnu = \left(0.37\pm0.06_{stat}\pm0.07_{syst}\right)\times 10^{-4}$.
The \Btoomegalnu branching-fraction measurement is the first with
a significance of more than five standard deviations.
It represents an improvement by a factor of
three over the only earlier measurement by Belle~\cite{belle}.
The improved measurements of \Btoomegalnu and \Btoetalnu decays
are important ingredients to the determination of the composition
of the inclusive charmless semileptonic decay rate.
The size of the data samples is not yet sufficient to perform
a measurement in intervals of the momentum transfer, $q^2$, of the decay,
which would be necessary to determine $|V_{ub}|$ with an adequate precision.

We are grateful for the excellent luminosity and machine conditions
provided by our \pep2\ colleagues, 
and for the substantial dedicated effort from
the computing organizations that support \babar.
The collaborating institutions wish to thank 
SLAC for its support and kind hospitality. 
This work is supported by
DOE
and NSF (USA),
NSERC (Canada),
IHEP (China),
CEA and
CNRS-IN2P3
(France),
BMBF and DFG
(Germany),
INFN (Italy),
FOM (The Netherlands),
NFR (Norway),
MIST (Russia), and
PPARC (United Kingdom). 
Individuals have received support from CONACyT (Mexico), A.~P.~Sloan Foundation, 
Research Corporation,
and Alexander von Humboldt Foundation.


\begin{thebibliography}{99}

\bibitem{CKM}
 M.~Kobayahi, T.~Maskawa,
 Prog. Theor. Phys. {\bf 49}, 652 (1973).

\bibitem{babarpilnu}
 \babar\ Collaboration, B.~Aubert \etal,
 Phys.~Rev.~Lett. {\bf 98}, 091801 (2007).

\bibitem{cleo}
CLEO Collaboration, S.B. Athar \etal,
Phys.~Rev.~{\bf D68}, 072003 (2003);
CLEO Collaboration, D. M. Asner \etal,
Phys.~Rev.~{\bf D76}, 012007 (2007).

\bibitem{babartagged}
 \babar\ Collaboration, B.~Aubert \etal, arXiv:hep-ex/0607066v1 
 (contribution to ICHEP, Moscow, 2006).

\bibitem{babarsltag}
 \babar\ Collaboration, B.~Aubert \etal, SLAC-PUB-13215, arXiv:0805.2408v1 [hep-ex] 
(submitted to Phys.~Rev.~Lett.).

\bibitem{belle}
 Belle Collaboration, C.~Schwanda \etal,
 Phys.~Rev.~Lett.~ {\bf 93}, 131803 (2004).

\bibitem{PDG}
W.-M. Yao \etal (Particle Data Group), 
J. Phys. {\bf G 33}, 1 (2006) and 2007 partial update for the 2008 edition.


\bibitem{babar} \babar\ Collaboration, B.~Aubert \etal,
 Nucl. Instrum. Methods {\bf A479}, 1 (2002).

\bibitem{lcsr:pi}
 P.~Ball, R.~Zwicky,
 Phys.~Rev.~{\bf D71}, 014015 (2005).

\bibitem{lcsr:rho}
 P.~Ball, R.~Zwicky,
 Phys.~Rev.~{\bf D71}, 014029 (2005).

\bibitem{lcsr:eta}
 P.~Ball, G.W.~Jones,
 JHEP {\bf 0708}, 025 (2007).

\bibitem{DFN}
 F.~De Fazio, M.~Neubert,
 JHEP {\bf 9906}, 017 (1999).

\bibitem{geant}
\geantfour Collaboration, S.~Agostinelli \etal,
Nucl. Instrum. Methods {\bf A506}, 250 (2003).

\bibitem{Vstar} All variables
denoted with a star (e.g. $p^*$) are given in the \FourS rest frame; all
others are in the laboratory frame except if differently specified.

\bibitem{R2} G. C. Fox and S. Wolfram, Phys. Rev. Lett. {\bf 41}, 1581 (1978).

\bibitem{L2} $L_2 =\sum_i |\vec{p}_i^{*}|\cos^2 (\theta_i^{*})$, where the sum is over all tracks in the event not used to form the $Y$ candidate and $\vec{p}_i^{*}$ and $\theta_i^{*}$ are their momenta and angles with respect to the thrust axis of the $Y$ candidate, respectively.

\bibitem{TMVA}
 A. H\"ocker \etal,
 arXiv:physics/0703039v4 [physics.data-an].

\bibitem{bbfit}
 R.~J. Barlow, C.~Beeston,
 Comput. Phys. Commun. {\bf 77}, 219 (1993).

\bibitem{DstarFF}
  \babar\ Collaboration, B.~Aubert \etal,  
  Phys.~Rev.~{\bf D77}, 032002 (2008). 
 
\bibitem{HFAG}
Heavy Flavor Averaging Group, E. Barberio \etal, ``Averages of b-hadron and c-hadron Properties at the End of 2007'', arXiv:0808.1297.

\bibitem{Bcounting}
  \babar\ Collaboration, B.~Aubert \etal,  
  Phys.~Rev.~{\bf D67}, 032002 (2003). 


\end{thebibliography}
\end{document}